\documentclass[twocolumn,showpacs,preprintnumbers,amsmath,amssymb]{revtex4}

\usepackage{graphicx}
\usepackage{dcolumn}
\usepackage{bm}

\begin{document}


\title{Dynamic relaxation of magnetic clusters in a ferromagnetic (Ga,Mn)As epilayer}

\author{K. Hamaya,$^{a)}$ T. Koike, T. Taniyama,$^{1}$ T. Fujii, Y. Kitamoto, and Y. Yamazaki}%
\affiliation{%
Department of Innovative and Engineered Materials, Tokyo Institute of
Technology,\\ 4259 Nagatsuta, Midori-ku, Yokohama 226-8502, Japan.\\
$^{1}$Materials and Structures Laboratory, Tokyo Institute of Technology,\\ 4259 Nagatsuta, Midori-ku, Yokohama 226-8503, Japan. 
}%


\date{\today}

\begin{abstract}
A new scenario of the mechanism of intriguing ferromagnetic properties in Mn-doped magnetic semiconductor (Ga,Mn)As is examined in detail. We find that magnetic features seen in zero-field cooled and field cooled magnetizations are not interpreted with a single domain model [Phys. Rev. Lett. {\bf 95}, 217204 (2005)], and the magnetic relaxation, which is similar to that seen in magnetic particles and granular systems, is becoming significant at temperatures above the lower-temperature peak in the temperature dependence of ac susceptibility, supporting the cluster/matrix model reported in our previous work [Phys. Rev. Lett. {\bf 94}, 147203 (2005)]. Cole-Cole analysis reveals that magnetic interactions between such (Ga,Mn)As clusters are significant at temperatures below the higher-temperature peak in the temperature dependent ac susceptibility. The magnetizations of these films disappear above the temperature showing the higher-temperature peak, which is generally referred to as the Curie temperature. However, we suggest that these combined results are evidence that the temperature is actually the blocking temperature of (Ga,Mn)As clusters with a relatively high hole concentration compared to the (Ga,Mn)As matrix.

\vspace{2mm}
$^{a)}$ Present adress: Institute of Industrial Science, University of Tokyo, 4-6-1 Komaba, Meguro-ku, Tokyo 153-8505, Japan. E-mail:  hamaya@iis.u-tokyo.ac.jp \\

\end{abstract}

\maketitle

\section{INTRODUCTION}
Although exchange interaction between Mn $d$ spins and hole carriers in the valence band likely causes ferromagnetism in Mn-doped magnetic semiconductor (Ga,Mn)As,\cite{Ohno2,Matsukura,Dietl2,Keavney} there are still controversies over the mechanism of the exotic magnetic features. One of the phenomena unpredictable in previous theories is the fact that magnetic uniaxial anisotropy along [110] occurs in the film plane and the easy axis changes from $\left\langle 100 \right\rangle$ to [110] with increasing temperature,\cite{Liu,Sawicki,Welp,Welp2,Hamaya2} which cannot be explained in terms of the hole induced mechanism. The second one is that first-principle calculations using a mean-field model estimate a rather high Curie temperature $T_{c}$ compared with experimental data.\cite{Sato} To address these issues, which are definitely vital to realization of ferromagnetism at room temperature, we have recently proposed a new possible mechanism of the switching of magnetic anisotropy on the basis of ac susceptibility measurements.\cite{HamayaPRL} The experimental results indicate that (Ga,Mn)As films on GaAs(001) consist of ferromagnetic (Ga,Mn)As cluster regions with [110] uniaxial anisotropy and ferromagnetic (Ga,Mn)As matrix with $\left\langle 100 \right\rangle$ cubic magnetocrystalline anisotropy, where the Curie temperature of the clusters is higher than that of the matrix due to relatively high hole concentration.\cite{HamayaPRL} 
Above the Curie temperature of the matrix, the magnetic moments of the clusters are even aligned along [110] due to their [110] uniaxial anisotropy, resulting that the magnetic easy axis switches from $\left\langle 100 \right\rangle$ to [110] near the Curie temperature of the matrix. 

In this paper, we show zero field cooled (ZFC) magnetization, field cooled (FC) magnetization, and ac susceptibility data of (Ga,Mn)As films grown on GaAs(001) more systematically. A clear bifurcation of the ZFC and FC magnetizations indicates the presence of inhomogeneous (Ga,Mn)As phases. Magnetization decay which is characteristic of an assembly of magnetic clusters is also clearly seen in the thermoremanent magnetization and the frequency dependent ac susceptibility above the temperature at which the magnetic easy axis switches from $\left\langle 100 \right\rangle$ to [110] in the time window of 10$^{-5}$ to 10$^{4}$ s. Analyses of these data indicate the presence of magnetic interaction between (Ga,Mn)As clusters. These dynamic magnetization relaxation data of a (Ga,Mn)As film ensure the validity of our cluster/matrix model.\cite{HamayaPRL} We also suggest that the temperature where the magnetization of (Ga,Mn)As disappears, which is generally referred to as the Curie temperature, is actually the blocking temperature of (Ga,Mn)As clusters embedded in (Ga,Mn)As matrix.  
\begin{figure}
\includegraphics[width=8.5cm]{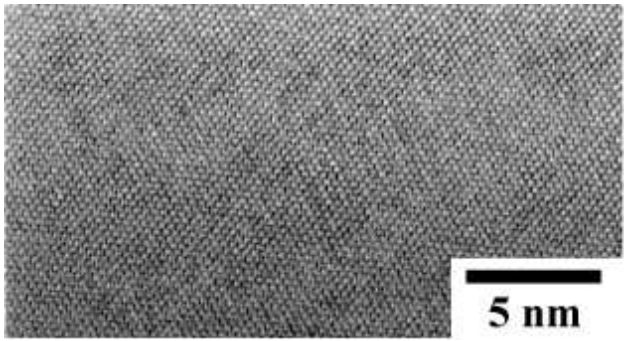}
\caption{HRTEM image of a (Ga,Mn)As epilayer.}
\end{figure}

\section{SAMPLE PREPARATION AND EXPERIMENTAL CONDITION}
In-plane magnetized (Ga,Mn)As films ($t=$100 nm, $T_{s}=$235$^{\circ}$C)/GaAs ($t=$400nm, $T_{s}=$590$^{\circ}$C) were grown on a semi-insulating GaAs (001) substrate using molecular beam epitaxy (MBE). Magnetic properties were measured using a superconducting quantum interference device (SQUID) magnetometer and a physical property measurement system (PPMS). We confirmed that the field dependent magnetizations ($M-H$ curves) of the as-grown sample at various temperatures showed magnetically anisotropic behavior, where switching of the magnetic easy axis from $\left\langle 100 \right\rangle$ to [110] occurs with increasing temperature as seen in literatures.\cite{Liu,Sawicki,Welp,Welp2,Hamaya2,Hamaya4,HamayaPRL} For ac susceptibility measurements, ac magnetic field with an amplitude of 5 Oe at various frequencies was applied parallel to $\left\langle 100 \right\rangle$ and in-phase $\chi'$ and out-of-phase $\chi''$ components of the linear ac susceptibility were recorded: $\chi$ $=$ $\chi'$ $+$ $i\chi''$.\cite{Mydosh} To rule out the possibility of NiAs-type MnAs precipitates, we checked cross-sectional transmission electron micrography (TEM) images of the film, ensuring  that there is no NiAs-type precipitates other than zinc-blende (Ga,Mn)As over the sample (Fig. 1). 
\begin{figure}[b]
\includegraphics[width=8.5cm]{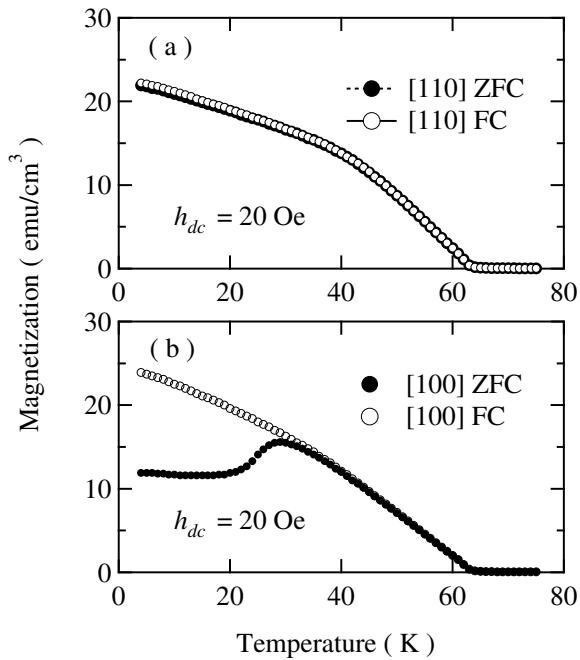}
\caption{Zero field cooled and field cooled magnetization as a function of temperature for (a) $h_{dc}$ $//$ [110] and (b) $h_{dc}$ $//$ [100] at an external field of 20 Oe.}
\end{figure}

\section{RESULTS AND ANALYSES}
\subsection{ZFC and FC dc magnetization behavior}
Figure 2 shows the temperature dependent magnetization in the ZFC and FC conditions at the field orientations of (a) [110] and (b) [100] at $h_{dc}$ $=$ 20 Oe. The ZFC curve follows the FC curve even in a small field of 20 Oe for [110] in Fig. 2(a). On the contrary, a clear bifurcation between the ZFC and FC magnetization curves is observed below 30 K for [100] in Fig. 2(b). A similar bifurcation behavior of ZFC and FC magnetization has been reported for other single crystal materials such as spin-glass or cluster-glass Fe$_{x}$TiS$_{2}$\cite{Koyano} and manganite perovskite La$_{1-x}$Ca$_{x}$MnO$_{3}$ with magnetic phase separation.\cite{Markovich} Since magnetic Mn ions are doped randomly in (Ga,Mn)As, we can infer that some inhomogeneous phases could exist in a (Ga,Mn)As single crystal in a similar manner. Therefore, the bifurcation of the ZFC and FC magnetizations may be attributed to randomly oriented magnetic domain structures in the ZFC condition.

\subsection{Ac susceptibility data and its Cole-Cole plot}
Figure 3 (a) shows the temperature dependence of remanent magnetization of an as-grown Ga$_{0.956}$Mn$_{0.044}$As film. The curves were recorded after field cooling in 1000 Oe at various field orientations. With increasing temperature, we can see switching of the magnetic easy axis from $\left\langle 100 \right\rangle$ to [110] at around 25 K.\cite{Sawicki,Welp,Hamaya2,HamayaPRL} The magnetization along the easy axis disappears at around 60 K which is regarded as the Curie temperature, i.e., $T_{c} \sim$ 60 K. For this sample, we also measure the temperature dependent ac susceptibility at various ac field orientations at a frequency of $\omega=$ 91 Hz in ac magnetic fields of $h_{ac}=$ 5, 6, 8 Oe and $h_{dc}$ $\sim$ 0 Oe. The $\chi'$$- T$ and $\chi''$$- T$ curves for the ac field orientation along $\left\langle 100 \right\rangle$ are shown in Fig. 3 (b). The dashed curves and solid curves denote $\chi'$ and $\chi''$, respectively. Two peaks are clearly seen in the temperature variation in Fig. 3(b) as given in our previous report.\cite{HamayaPRL} We define the lower temperature and the higher temperature corresponding to the two peaks as $T_{L}$ and $T_{H}$, respectively. $T_{H}$ is most likely to be associated with $T_{c} \sim$ 60 K of this sample although we could not detect signals at smaller ac fields in our equipment. In our view, these two peaks originate from inhomogeneous mixed magnetic phases with two different magnetic anisotropies.\cite{HamayaPRL} As ac magnetic field is increased, the ac susceptibility is enhanced, and both $T_{L}$ and $T_{H}$ shift toward lower temperatures. We also show $\chi'$$- T$ and $\chi''$$- T$ curves at the ac field orientations along [110] (Fig. 3(c)) and [1$\overline{1}$0] (Fig. 3(d)) in the same measurement conditions. A peak of the ac susceptibility is seen at $T_{H}$, while that at $T_{L}$ disappears completely for the ac field orientation along [110] in Fig. 3(c). For the ac field along [1$\overline{1}$0], on the other hand, the ac susceptibility shows a large peak at $T_{L}$ accompanied with a very small peak at $T_{H}$. This ac field orientation dependence of the ac susceptibility is observed for other several samples we grew. Very recently, similar results have been reported by Wang {\it et al}.\cite{Wang}
\begin{figure}
\includegraphics[width=8.5cm]{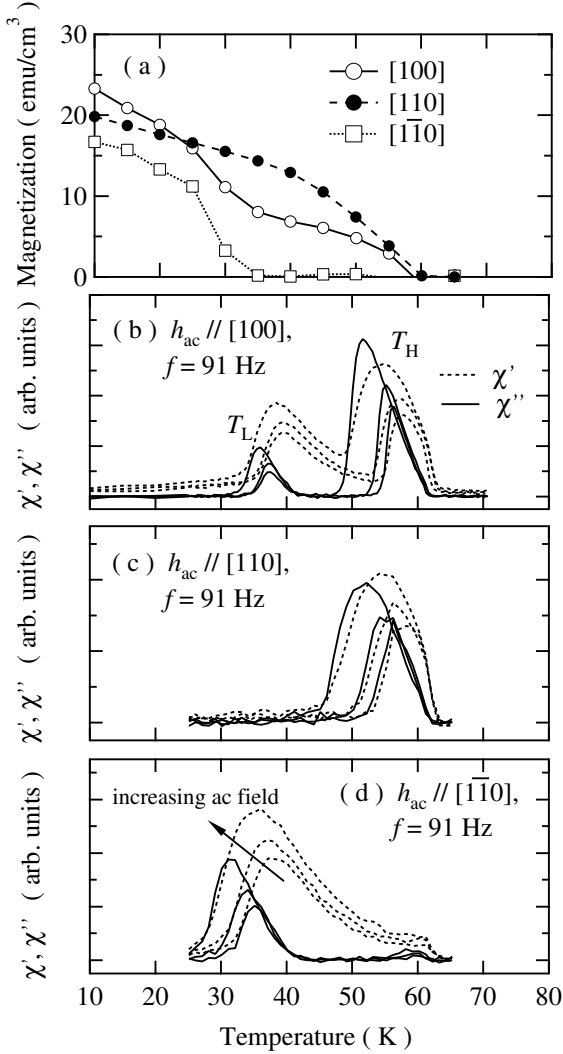}
\caption{(a) Temperature dependent remanent magnetization of an as-grown sample for various field orientations. Out-of-phase components of ac susceptibility $\chi''$ as a function of temperature for ac field orientations of (b) $\left\langle 100 \right\rangle$ (c) [110] and (d)  [1$\overline{1}$0] at a frequency of $\omega=$ 91 Hz in ac magnetic fields of $h_{ac}=$ 5, 6, 8 Oe.}
\end{figure}

Shown in Fig. 4(a) is Cole-Cole plots\cite{ColeCole} of the ac susceptibility for $h_{ac}$ // $\left\langle 100 \right\rangle$ at around $T_{H}$, constructed from the frequency dependence of the ac susceptibilities. Since the Cole-Cole plots have the shape of an arc, the dynamic behavior of a (Ga,Mn)As film can be represented in terms of Debye-type relaxation with a distribution of the relaxation time. To examine the relaxation function in detail, we show the frequency dependence of $\chi''$, i.e., $\chi'' - f$ at around $T_{H}$ in Fig. 4(b) since $\chi''$ is proportional to the distribution function of the relaxation time $g$($\tau$).\cite{Lundgren,Mydosh} In general, complex susceptibility is expressed as \cite{Ravindran,Dekker} 
\begin{equation}
\chi = \chi_{s} + \frac{\chi_{0} - \chi_{s}}{1 + (i\omega\tau_{c})^{1-\alpha}},
\end{equation}
where $\chi_{0}$ and $\chi_{s}$ are the isothermal ($\omega =$ 0) and adiabatic ($\omega$ $\rightarrow$ $\infty$) susceptibilities, respectively, $\tau_{c}$ is the median relaxation time of a distribution, and $\alpha$ (0$<$$\alpha$$<$1) is associated with the width of the distribution. 
Dividing  Eq.(1) into the in-phase and out-of-phase components, $\chi''$ is obtained as follows:\cite{Ravindran,Dekker}  
\begin{equation}
\chi'' = \frac{\chi_{0} - \chi_{s}}{2} \frac{cos(\pi\alpha/2)}{cosh[(1-\alpha)ln(\omega\tau_{c})]+sin(\pi\alpha/2)}, 
\end{equation}
where $\omega_{c}$ ($=$ 1/$\tau_{c}$) is the characteristic frequency corresponding to the maximum of $\chi''$. We fit Eq. (2) to the experimental data in Fig. 4(b) and the best fit curves are also shown in Fig. 4(b), indicating that the characteristic magnetization relaxation time $\tau_{c}$ becomes short as the temperature increases near $T_{H}$. 
\begin{figure}
\includegraphics[width=8.5cm]{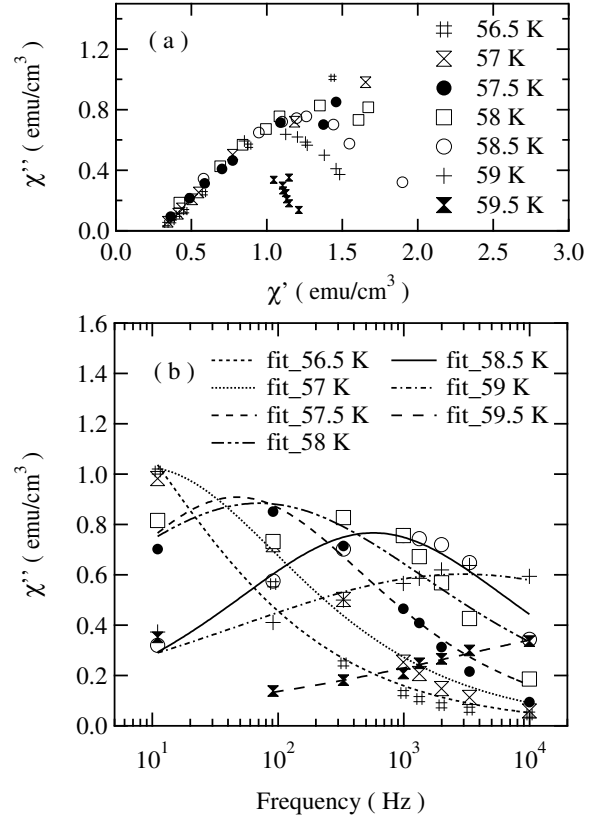}
\caption{(a) Cole-Cole plots of the ac susceptibilities at around $T_{H}$. (b) Frequency dependence of $\chi''$ and the best fit curves using Eq. (2) are shown.}
\end{figure}
 
 \subsection{Long time relaxation of thermoremanent magnetization}
We hereafter describe the dynamic magnetization behavior of as-grown (Ga,Mn)As. Figure 5 shows semilog plots of the time decay of thermoremanent magnetization (TRM) after field cooling in $h_{dc} =$ 20 Oe along $\left\langle 100 \right\rangle$ down to a measurement temperature. The values are normalized with the magnetization at $t =$ 0 s, where we define $t =$ 0 s as the time when the first data point is recorded. A clear decay of the TRM is observed over the time window of 10$^{3}$ s above 40 K, whereas no decay can be seen at 10 K. Similar relaxation was observed in the anomalous Hall resistivity of ferromagnetic semiconductor (In,Mn)As/GaSb heterostructures,\cite{Oiwa6} 
although the particular decay form of (Ga,Mn)As at 50 K is different from those of (In,Mn)As/GaSb. 
It is also well known that such characteristic logarithmic magnetization relaxation is observed in dispersed magnetic particle systems, diluted magnetic alloys, and magnetic granular films.\cite{Thermo} Following the analyses used in these studies, we analyze the time decay of the magnetization using a stretched exponential relaxation at 40, 45, and 50 K, as given by \cite{Mydosh,Thermo}
\begin{equation}
M_\text{TRM} = M_{0} exp[-(\omega t)^{\beta}], (0 < \beta < 1),
\end{equation}
where $M_\text{TRM}$ is the TRM, $\omega$ is the relaxation frequency, $M$$_{0}$ and $\beta$ are the temperature dependent constants, and $\beta =$ 1 corresponds to the Debye-type relaxation. The inset of Fig. 5 presents both the raw TRM data at 50 K and a fitted curve using Eq. (3). The magnetization relaxation time of the (Ga,Mn)As film above $T_{L}$, which is estimated using Eq. (1) and Eq. (3), are summarized in Fig. 6 (a). With increasing temperature, a significant decrease in $\tau_{c}$ from $\sim$10$^{12}$ to $\sim$10$^{-6}$ s is observed, being similar to a marked reduction in the relaxation time of cluster-glass materials.\cite{Koyano} Therefore, the dynamic relaxation behaviors in (Ga,Mn)As are likely arising from thermally activated magnetization reversal of clusters in the (Ga,Mn)As film as seen in monodispersed particles systems and granular systems\cite{Huser} such as Fe-C,\cite{Djurberg} Fe$_{2}$O$_{3}$,\cite{Poddar} or Co:Al$_{2}$O$_{3}$\cite{Luis}, which is consistent with the cluster/matrix model described in our previous work.\cite{HamayaPRL} 
\begin{figure}
\includegraphics[width=8cm]{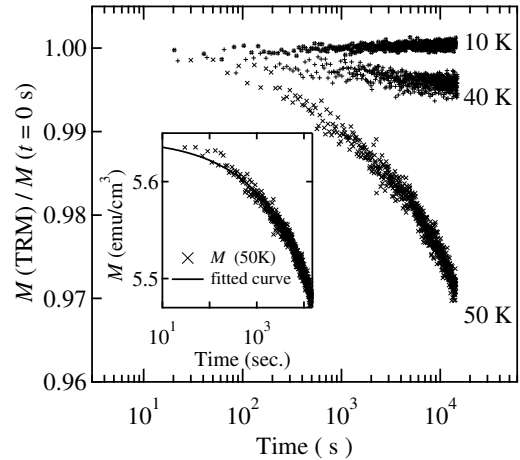}
\caption{Normalized time decay curves of the thermoremanent magnetization at various temperatures. The inset shows the raw data at 50 K and a fit using a stretched exponential curve in Eq. (3).}
\end{figure}

\subsection{Annealing effect on dc and ac magnetization behavior}
Since low temperature annealing at 200$^{\circ}$C is predicted to dissociate pairs of interstitial Mn and substitutional Mn in an as-grown sample, the annealing process accelerates an increase in the hole concentration up to $\sim$10$^{21}$cm$^{-3}$ and consequently improves the magnetic quality of (Ga,Mn)As as described by Edmonds {\it et al}.\cite{Edmonds} To examine the relationship between dc magnetization and ac susceptibility for annealed samples, we show dc and ac magnetization of a Ga$_{0.956}$Mn$_{0.044}$As film annealed for 16 hours in Figs. 7. Figure 7(a) depicts the remanent magnetization as a function of temperature after field cooling at various field orientation in $h_{dc}$ $=$ 1000 Oe. As clearly seen, the [110] uniaxial anisotropy dominates the magnetic behavior over all the temperature range for this 16 hour-annealed sample, being compatible with a recent study by Sawicki {\it et al}.\cite{Sawicki2} 
In Fig. 7(a), we also show $M_{[110]}\cos45^{\circ}$ which is the [100] component of the [110] remanent magnetization (solid curve). The values of $M_{[110]}\cos45^{\circ}$ well agree with the remanent magnetization along [100], $M_{[100]}$, above 35 K, while the deviation between $M_{[100]}$ and $M_{[110]}\cos45^{\circ}$ is significant below 35 K, indicating the presence of inhomogeneous phases even for the long time annealed sample as stated in the previous sections.

The $\chi''$$- T$ curves of this sample are shown in Fig. 7(b) at various ac field orientations. All the curves have a very broad single peak. It should be noted that a peak in the $\chi''$$- T$ appears at around 40 K which corresponds to the $T_{L}$ of this sample despite the fact that no switching of the magnetic easy axis occurs at the ac field orientation along [1$\overline{1}$0] in Fig. 7(a). The results cannot be understood with reorientational switching of the magnetic anisotropy at the $T_{L}$.
\begin{figure}
\includegraphics[width=8.5cm]{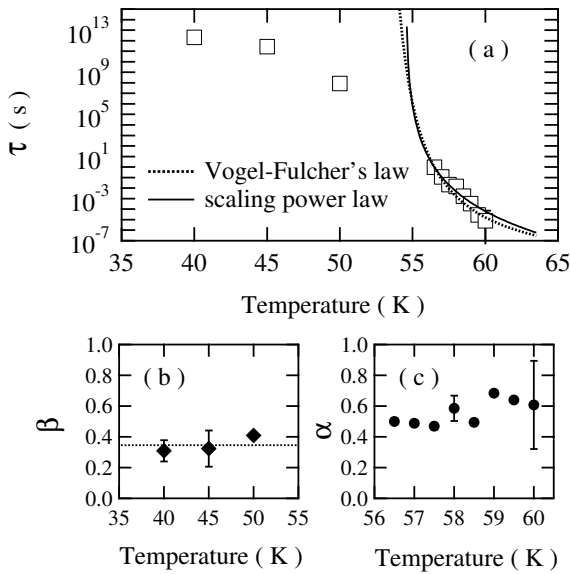}
\caption{(a) Temperature dependent relaxation time above $T_{L}$, estimated from fitting using Eq. (2) and Eq. (3). (b) Temperature dependence of $\beta$ in the temperature range 40 K $\lesssim$ $T$ $\lesssim$ 50 K. (c) Temperature dependence of $\alpha$ at around $T_{H}$.}
\end{figure}

\section{DISCUSSION}
\subsection{Magnetization reorientation model vs. cluster/matrix model}
Recently, Wang {\it et al}. pointed out that the field orientation dependence of the ac susceptibility shown in Fig. 3 is fully understood using a magnetization reorientation model of single domain (Ga,Mn)As.\cite{Wang} As far as the data shown in Fig. 3 are concerned, their interpretation seems correct. However, we have shown other magnetization results which are incompatible with the single domain scenario. As described in the previous section, the bifurcation of the ZFC and FC magnetization curves is seen only for the [100] field orientation in Fig. 2(b). If we simply take a single domain structure with cubic magnetic anisotropy into account, the ZFC curve should follow the FC curve irrespective of field orientations. Thus, we need alternative description which explains the bifurcation of ZFC and FC magnetization - cluster/matrix model.

In general, (Ga,Mn)As with a high hole concentration is believed to have [110] uniaxial anisotropy and (Ga,Mn)As clusters in the cluster/matrix model likely have [110] magnetic easy axis.\cite{HamayaPRL} When such a sample is cooled down to a measurement temperature in zero field, the initial magnetic orientation of the clusters at the low temperature is completely random. Once a small magnetic field is applied along [110], the magnetization of the clusters is immediately aligned along [110] due to the intrinsic [110] anisotropy of the clusters. On the other hand, if magnetic field is applied along [100] after zero field cooling, the magnetization of clusters is hard to be aligned along [100], resulting in the bifurcation of ZFC and FC magnetizations only along [100]. Thus this description based on our previous interpretation explains the experimental data comprehensively.\cite{HamayaPRL} 
\begin{figure}
\includegraphics[width=8.5cm]{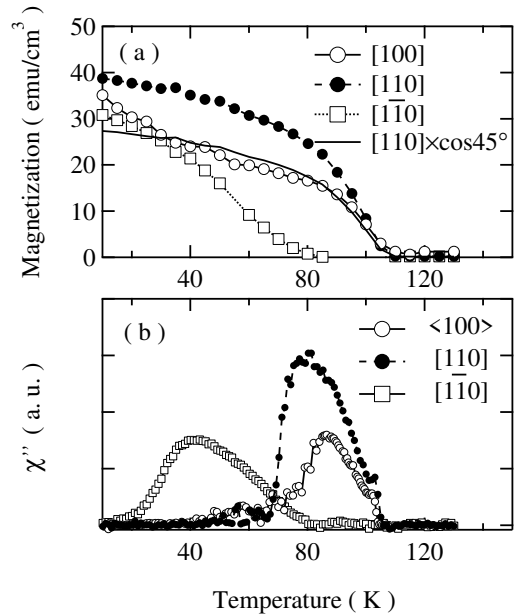}
\caption{(a) Temperature dependent remanent magnetization of a 16-hour-annealed sample. (b) Out-of-phase component of ac susceptibility $\chi''$ as a function of temperature for various ac field orientations.}
\end{figure}

Moreover, we have found that the ac susceptibility of a long time annealed sample shows a peak at  $T_{L}$ for the ac field orientation along [1$\overline{1}$0] nevertheless no switching of the magnetic easy axis in Fig. 7(a). Also, the values of $M_{[110]}\cos45^{\circ}$ deviate from $M_{[100]}$ below $T_{L}$. If we assume the cluster/matrix model again, the no switching behavior of the magnetic easy axis can be understood through the growth of the cluster phase with annealing and the consequent decrease in the magnetic inhomogeneity. In other words, the cluster region with [110] uniaxial anisotropy in an as-grown sample becomes dominant after annealing, while the volume of the matrix phase with $\left\langle 100 \right\rangle$ cubic anisotropy in the as-grown sample decreases, resulting in another mixed phase in which the majority of the sample are composed of the phase with [110] uniaxial anisotropy.\cite{Hamaya4}

Edmonds {\it et al.} studied annealing effect on the magnetic properties of (Ga,Mn)As, recently, suggesting the presence of inhomogeneous distribution of doped Mn concentration.\cite{Edmonds} Also, Sawicki {\it et al.} have reported an enhancement in the [110] uniaxial anisotropy for samples subjected to low temperature annealing due to a small trigonal-like distortion associated with inhomogeneous Mn distribution.\cite{Sawicki2} These information about the inhomogeneous Mn distribution may also be associated with the formation of [110] (Ga,Mn)As clusters, and thus the inhomogeneous features are unlikely arising from a single phase model proposed by Wang {\it et al}.\cite{Wang} Therefore, the cluster /matrix model is more realistic for understanding microscopic magnetic properties of as-grown (Ga,Mn)As rather than the single phase magnetic structure.

\subsection{Origin of characteristic magnetic relaxation}
From the analysis of $\tau_{c}$, we classify the characteristic relaxation behavior into (1) that at 40 K $\lesssim$ $T$ $\lesssim$ 50 K and (2) that at $T$ $\approx$ $T_{H}$. In the regime (1), stretched exponential decay of the TRM in Fig. 5 clearly shows the existence of magnetic interaction between (Ga,Mn)As clusters in the epilayer. For an assembly of magnetic particles, there are a few possible mechanisms of magnetic interaction between particles.\cite{Batllle} In general, dipole-dipole interactions between particles are primarily important for such systems. Direct exchange interaction via the surface of particles are also present, given the particles are in close contact with each other. However, the content of clusters in the (Ga,Mn)As film is not significant so this contribution to the magnetic interaction between particles is almost negligible.\cite{foot} If particles are embedded in a metallic matrix, RKKY interaction may be expected. This might be the case for (Ga,Mn)As, since (Ga,Mn)As clusters are embedded in the matrix with hole carriers which mediate the interaction between the clusters.\cite{HamayaPRL} As shown in Fig. 6(b), the constant value of $\beta$ which is independent of temperature in this temperature regime, i.e., $T$/$T_{H}$ $<$ 0.8, is also compatible with the description of magnetic interaction as seen in metallic spin-glass \cite{Chamberlin} and an assembly of magnetic particles.

Given a non-interacting clusters assembly in the temperature regime (2), on the other hand, the long time relaxation of the magnetization could be represented by a N\'eel-Brown model, i.e., $\tau$ $=$ $\tau_{0}$exp($E$$_{a}$/$k_{B}$$T$),\cite{NB} where $\tau_{0}$ is the inverse of the frequency of a cluster to reverse the magnetization, $E$$_{a}$ is the anisotropy energy, and $k_{B}$ is the Boltzmann's constant. Although we have made an attempt to fit the N\'eel-Brown model to the experimental data near $T_{H}$, the fitting is rather poor, giving an unphysical value of $\tau_{0}$ $\ll$ 10$^{-14}$ s. Therefore, even at $T$ $\approx$ $T_{H}$, magnetic interaction between (Ga,Mn)As clusters is significant. 

According to previous works on spin-glass and interacting monodispersed particles systems,\cite{Mydosh,Luis,Denardin,Djurberg,Poddar} critical slowing down of the relaxation behavior can be fitted either by a modified exponential law (Vogel-Fulcher's law), i.e., $\tau$ $=$ $\tau_{0}$exp($E_{a}$/$k_{B}$($T-T_{0}$)) or by a scaling power law, i.e., $\tau$ $=$ $\tau$*($T$/$T_{b}$$-$1)$^{-z\nu}$, $T$ $>$ $T_{b}$, where $T_{0}$ is an effective temperature which accounts for the interaction effect, $T_{b}$ is the transition temperature, $\tau$* is related to the relaxation time of the magnetic moments of individual clusters, and $z$$\nu$ is the critical exponent composed of a static part $\nu$ and a dynamic part $z$. We first fit the Vogel-Fulcher's law to the experimental data near $T_{H}$ as shown in Fig. 6 (a) (dashed line), giving $T_{0} =$ 52.5 K, $E$$_{a}$/$k_{B}$ $=$ 94.7 K, and $\tau_{0}$ $=$ 5.6 $\times$ 10$^{-11}$ s. The values of these parameters are in good agreement with those of a diluted and monodispersed interacting Fe particle system embeded in C matrix.\cite{Djurberg} On the other hand, a fit to the experimental data using the scaling power law provides $\tau$* $=$ 2.3 $\times$ 10$^{-14}$ s, $T_{b}$ $=$ 54.5 K and $z$$\nu$ $=$ 9.45 $\pm$ 3. Although the values of $\tau$* and $z$$\nu$ are very close to those for insulating spin-glass Cd$_{0.6}$Mn$_{0.4}$Te,\cite{Mauger} the mechanism of spin freezing in Cd$_{0.6}$Mn$_{0.4}$Te, which is the frustration of antiferromagnetically superexchange coupled spins, is totally different from that of (Ga,Mn)As as described before. Therefore we conclude that magnetic relaxation near $T_{H}$ are similar to those of the diluted and monodispersed Fe/C system,\cite{Djurberg} and hole-mediated exchange interaction or the dipole-dipole interaction between clusters primarily explains the Vogel-Fulcher's relaxation behavior. 

A fit of Eq. (2) to the $\chi''$$- f$ curves in Fig. 4(b) provides the temperature dependence of the parameter $\alpha$ near $T_{H}$ (Fig. 6(c)). The $\alpha$ is known to be associated with the width of the distribution of $\tau_{c}$: $\alpha =$ 1 corresponds to an infinitely wide distribution while $\alpha =$ 0 represents the Debye type relaxation. In general, spin-glass and cluster-glass materials have a wide distribution of $\tau_{c}$ so that the $\alpha$ deviates from $\alpha =1$ and decreases with increasing temperature.\cite{Mydosh,Koyano,FMatsukura} Strongly interacting particles systems also show such behavior with increasing temperature.\cite{Huser} However, for (Ga,Mn)As the $\alpha$ has no systematic change as a function of temperature as shown in Fig. 6(c) although we regard (Ga,Mn)As as an interacting cluster system. The discrepancy between $\alpha$ obtained in this study and the general feature may be associated with the very narrow temperature range from 56.5 K to 60.0 K used in this study. So provided that a wider temperature range is considered, a clearer temperature dependence of $\alpha$ could be obtained. We note, however, that low temperature annealing which decreases local defects such as Mn interstitials does not change the feature near $T_{H}$ qualitatively, while the peak in the temperature dependent ac susceptibility at $T_{H}$ shifts toward higher temperatures.\cite{Hamaya4}

The discussion about the magnetic relaxation in (Ga,Mn)As provides the following comprehensive description for the controversy over the magnetic anisotropy and magnetization problem in view of our cluster-matrix model. The magnetic relaxation in (Ga,Mn)As clearly resembles those of magnetic particles assemblies and magnetic granular systems above $T_{L}$, indicating that (Ga,Mn)As clusters, which have a relatively high hole concentration, dominate the relaxation behavior. As the temperature reaches $T_{H}$, the magnetization of the clusters begins to melt due to thermal fluctuation since the magnetic anisotropy of the clusters is very small.\cite{HamayaPRL} Mn moments inside the cluster, however, are likely still aligned via $p-d$ or double exchange interaction at $T_{H}$, that is, the temperature where the magnetization disappears in the $M-T$ curve, which is generally referred to as the Curie temperature, is actually the blocking temperature of the (Ga,Mn)As clusters. Therefore, we suggest that magnetization measurements carried out so far have probed a blocking temperature as the Curie temperature while first-principle calculation \cite{Sato} estimates the Curie temperature where the magnetic alignment of individual Mn moments inside the clusters disappears - the Curie temperature is higher than the blocking temperature in general. This is the primary reason of the discrepancy between experiments and theoretical calculation reported.

\section{CONCLUSION}
We have discussed the results of zero field cooled (ZFC) and field cooled (FC) magnetizations and the dynamic magnetic relaxation in (Ga,Mn)As films to ensure the validity of our cluster/matrix model. The bifurcation between ZFC and FC curves cannot be interpreted using a single domain model of (Ga,Mn)As with a single magnetic phase. Magnetization relaxation behavior similar to that of a magnetic particles assembly and magnetic granular systems is clearly observed above $T_{L}$, where the magnetic easy axis switches from $\left\langle 100 \right\rangle$ to [110], in the Cole-Cole analyses. The characteristic dynamic relaxation can be explained in terms of magnetic interaction between (Ga,Mn)As clusters with a relatively high hole concentration. Also, we have proposed that the temperature where the macroscopic magnetization disappears is not the Curie temperature of (Ga,Mn)As but the blocking temperature of the highly hole concentrated (Ga,Mn)As clusters. 

 \begin{acknowledgments}
We acknowledge Prof. Hiro Munekata of Tokyo Institute of Technology for providing (Ga,Mn)As samples used in this study. 
K. H. acknowledges support from the 21st Century COE Program " Nanomaterial Frontier Cultivation for Industrial Collaboration" of MEXT, Japan.
\end{acknowledgments}
\vspace{5mm}



\begin{thebibliography}{11}
\bibitem{foot}
{(Ga,Mn)As clusters are embedded in a conductive (Ga,Mn)As matrix, and we assume that these clusters are not in contact with each other in this diluted magnetic system.}
\end{thebibliography}
\end{document}